\documentclass[12pt]{article}

\newcommand{\bm}[1]{{\boldsymbol {#1}}}
\newcommand{\diag}{\,\mathrm{diag}\,}%

\usepackage{fullpage}

\usepackage{times}
\usepackage{bm}
\usepackage{natbib}
\usepackage{amsmath, amsbsy, amssymb, booktabs, amsmath, graphicx, subfigure}

\usepackage{ifpdf}

\ifpdf
\usepackage[pdftex]{hyperref}
\else
\usepackage[hypertex]{hyperref}
\fi

\usepackage{authblk, url}

\usepackage[pagewise]{lineno}
\linenumbers*[1]
\newcommand*\patchAmsMathEnvironmentForLineno[1]{%
  \expandafter\let\csname old#1\expandafter\endcsname\csname #1\endcsname
  \expandafter\let\csname oldend#1\expandafter\endcsname\csname end#1\endcsname
  \renewenvironment{#1}%
  {\linenomath\csname old#1\endcsname}%
  {\csname oldend#1\endcsname\endlinenomath}}%
\newcommand*\patchBothAmsMathEnvironmentsForLineno[1]{%
  \patchAmsMathEnvironmentForLineno{#1}%
  \patchAmsMathEnvironmentForLineno{#1*}}%
\AtBeginDocument{%
  \patchBothAmsMathEnvironmentsForLineno{equation}%
  \patchBothAmsMathEnvironmentsForLineno{align}%
  \patchBothAmsMathEnvironmentsForLineno{flalign}%
  \patchBothAmsMathEnvironmentsForLineno{alignat}%
  \patchBothAmsMathEnvironmentsForLineno{gather}%
  \patchBothAmsMathEnvironmentsForLineno{multline}%
}

\newcommand\email[1]{#1}
\def\keywordfont{\fontsize{9}{10}\selectfont}
\newenvironment{keywords}{\par\addvspace{8pt}%
                          \keywordfont\noindent{\itshape Some key words}:\ \ignorespaces%
                         }{\par\addvspace{23pt}}
\graphicspath{{.:./figs/}}

\begin{document}

\title{Transformed Gaussian Markov Random Fields and Spatial Modeling}

\author[1]{Marcos O. Prates}
\affil[1]{Departamento de Estat\'{i}stica, Universidade Federal de Minas Gerais,
Av. Presidente Ant\^{o}nio Carlos 6627 Pampulha Belo Horizonte, Minas Gerais, 31270-901, Brazil \email{marcosop@est.ufmg.br} }

\author[2]{Dipak K. Dey}
\affil[2]{Department of Statistics, University of Connecticut, 215 Glenbrook Rd. U-4120 Storrs, Connecticut 06269, U.S.A. \email{dipak.dey@uconn.edu} }

\author[3]{Michael R. Willig}
\affil[3]{Department of Ecology \& Evolutionary Biology, University of Connecticut, 75 N. Eagleville Rd. U-3043
Storrs, Connecticut 06269, U.S.A. \email{michael.willig@uconn.edu}}

\author[2]{Jun Yan}
\affil{Department of Statistics, University of Connecticut, 215 Glenbrook Rd. U-4120
Storrs, Connecticut 06269, U.S.A. \email{jun.yan@uconn.edu}}

\maketitle

\begin{abstract}
The Gaussian random field (GRF) and the Gaussian Markov random
field (GMRF) have been widely used to accommodate spatial
dependence under the generalized linear mixed model framework.
These models have limitations rooted
in the symmetry and thin tail of the Gaussian distribution.
We introduce a new class of random fields,
termed transformed GRF (TGRF), and a new class of Markov
random fields, termed transformed GMRF (TGMRF).
They are constructed by transforming the margins of GRFs and
GMRFs, respectively, to desired marginal distributions to
accommodate asymmetry and heavy tail as needed in practice.
The Gaussian copula that characterizes the dependence structure
facilitates inferences and applications in modeling spatial dependence.
This construction leads to new models such as gamma or beta Markov 
fields with Gaussian copulas, which can be used to model Poisson 
intensity or Bernoulli rate in a spatial generalized linear mixed model.
The method is naturally implemented in a Bayesian framework.
We illustrate the utility of the methodology in an ecological
application with spatial count data and spatial presence/absence
data of some snail species, where the new models are
shown to outperform the traditional spatial models.
The validity of Bayesian inferences and model selection
are assessed through simulation studies for both spatial
Poisson regression and spatial Bernoulli regression.

\end{abstract}


\section{Introduction}
\label{s:intro}

A Gaussian random field (GRF) is a stochastic process whose finite 
dimensional marginal distribution of any dimension is Gaussian.
A GRF that enjoys the Markov property is a Gaussian Markov
random field (GMRF) and can be represented by an undirected graph.
Specifically, a random vector $\bm{Z}$ follows a GMRF with respect
to some labeled graph $\mathcal{G}$ if $\bm{Z}$ is a GRF with
precision matrix $\bm{Q}$ such that $Q_{ij} \neq 0$ if and only if
$i$ and $j$ are connected in graph $\mathcal{G}$ \citep{rue:held}.
A GMRF model is natural when specification of precision $\bm{Q}$ 
is easier than specification of covariance, $\bm{\Sigma}= \bm{Q}^{-1}$.
From the Markov property, the conditional distribution of node $i$ in
$\mathcal{G}$ given all the rest is fully specified by its neighbors.
The precision matrix $\bm{Q}$ plays an essential role here and the
sparsity of $\bm{Q}$ facilitates fast sampling algorithms, which
is important in sampling based inferences \citep{Rue:2001}.
GMRFs are analytically tractable, can be easily used in hierarchical
models, and fit nicely in a Bayesian framework.

The GMRF model has been widely used in a variety of fields.
Along with the public concerns in the environment and public health,
recent applications have surged in environmental sciences
\citep[e.g.,][]{Wikle:etal:1998, Huerta:etal:2004,
Rue:etal:2004} and epidemiology \citep[e.g.,][]{besag:york:mollie,
Knorr-Held:Besag:1998, Knorr-Held:Rue:2002, Schmid:Held:2004}.
In particular, GMRFs are used as random effects to account for 
spatial dependence in the generalized linear mixed model (GLMM) 
framework; see \citet{rue:held} and references therein.
Popular as they are, GMRFs are limited in accommodating asymmetry
or heavy tails in practice because of the marginal Gaussian property.
To the best of our knowledge, Markov random fields with arbitrary margins
are underdeveloped for modeling data with asymmetry or heavy tails.

We propose random field that are transformed from a GRF and GMRF
such that the marginal distributions are of any desired form,
through the probability integral transformation and its inverse.
By Sklar's theorem \citep{Sklar:1959}, any continuous multivariate 
distribution can be uniquely represented by its marginal distributions 
and a copula which characterizes the dependence structure.
Since copulas are invariant to monotonic transformations,
the dependence structure of the new field, in terms of
copula, remains the same as that of the GRF or GMRF.
Such construction leads to new random fields with desired margins
combined with Gaussian copulas, such as gamma fields or beta fields,
which may be used to model spatial Poisson intensity or Bernoulli rate.
The transformed GRF (TGRF) and the transformed GMRF (TGMRF)
share some of the properties of the GRF and GMRF, respectively.
Due to the marginal transformations, they allow a general and flexible 
representation that can easily accommodate asymmetry as well as heavy 
tail behavior that are often observed in empirical data.

The rest of the article is organized as follows.
In Section~\ref{s:RanField}, TGRFs and TGMRFs are defined
and some of their important properties are presented.
In Section~\ref{s:glmm}, TGMRFs are incorporated into
a GLMM framework to accommodate spatial dependence.
In Section~\ref{s:po} and Section~\ref{s:bin}, the proposed
models are applied to the count data and presence/absence
data, respectively, of some snail species in an ecological study.
The statistical inferences are made in the Bayesian framework.
The performance of the inferences and model selection in both
applications are investigated in simulations that mimic the real data.
A discussion concludes in Section~\ref{s:conc}.

\section{A General Class of Random Fields}
\label{s:RanField}

For ease of notation, we present the definition of a TGRF
and a TGMRF in the context of finite dimension $n$ in the sequel.
For random fields indexed by elements in some space, the definition
applies to $n$-dimensional marginal distributions for any $n$.

% \subsection{Transformed Gaussian Random Field}

Suppose that $\bm{\varepsilon} = (\varepsilon_1,\ldots,\varepsilon_n)^{\top}$
is $n$-dimensional standard multivariate normal with mean $\bm{0}$ and
correlation matrix, $\bm{\Psi}$, denoted as $N_n(\bm{0},\bm{\Psi})$.
Define a random vector $\bm{Z} = (Z_1,\ldots,Z_n)^{\top}$ through
\begin{equation}
\label{eq:Z}
Z_i = F_{i}^{-1}\left\{\Phi\left(\varepsilon_i)\right)\right\},
\quad i = 1, \ldots, n,
\end{equation}
where $F_{i}$ is the distribution function of an absolutely continuous
variable and $\Phi$ is the distribution function of $N(0,1)$.
Then, each $Z_i$ has a marginal distribution $F_{i}$.
The random vector $\bm{Z}$ is called a TGRF with symmetric positive
definite (s.p.d.) dependence matrix $\bm{\Psi}$, denoted as
$\mathrm{TGRF}_n(\bm{F},\bm{\Psi})$, where $\bm{F} = (F_{1},\ldots,F_{n})$.
The joint density of $\bm{Z}$ can be easily shown to be
\begin{equation}
    h(\bm{x}) = (2\pi)^{-\frac{n}{2}} |\bm{\Psi}|^{-\frac{1}{2}} \exp\left(-\frac{1}{2} \bm{\varepsilon}^{\top} \bm{\Psi}^{-1} \bm{\varepsilon}\right) \prod_{i=1}^n \frac{f_i(x_i;\bm{\theta}_i)}{\phi(\varepsilon_i)},
\label{eq:mvgPdf}
\end{equation}
where $f_i$ is the density corresponding to $F_{i}$ with parameters
$\bm{\theta}_i$, $i = 1, \ldots, n$, $\phi$ is the density of $N(0,1)$,
and $\bm{\varepsilon}=\Big[\Phi^{-1}\big\{\mathrm{F}_1(x_1)\big\} ,\ldots,\Phi^{-1}\big\{\mathrm{F}_n(x_n)\big\}\Big]^\top$.

Clearly, a TGRF is obtained by transforming all the margins of a GRF
with standard normal margins to desired marginal distributions $F_i$'s.
The resulting TGRF is not affected by the scales of the original GRF
since we can always standardize the margins to standard normals.
The dependence structure of the TGRF, its copula, is still the Gaussian 
copula of the original GRF \citep[e.g.,][]{Joe:1997, Nelsen:2006}.
It is characterized by matrix $\bm{\Psi}$, but $\bm{\Psi}$ 
no longer has the interpretation of correlation matrix.
A $\mathrm{TGRF}_n(\bm{F},\bm{\Psi})$ is completely specified by
marginal distributions $\bm{F}$ and a Gaussian copula specified by a
dispersion matrix $\bm{\Psi}$. 
More details and properties of the TGRF are presented in 2011 
University of Connecticut PhD thesis by M.~O.~Prates.
% \citet{Prates:2011}.

A $\mathrm{TGRF}_n(\bm{F}, \bm{\Psi})$ is a TGMRF if the GRF before the
transformations is a standard GMRF with correlation matrix $\bm{\Psi}$.
As commonly used for GMRFs, it is more convenient to present a TGMRF
using the precision matrix $\bm{Q} = \bm{\Psi}^{-1}$, since it leads to
an intuitive interpretation of conditional distributional properties.
Let $\bm{Z}$ be a $\mathrm{TGRF}_n(\bm{F}, \bm{Q}^{-1})$
with a s.p.d. pre-transformation precision matrix $\bm{Q}$. 
Since the transformations are marginal-wise, the Markov property 
is inherited by the TGMRF: for $i \neq j$,
$Z_i \perp Z_j | \bm{Z}_{(-ij)}$ if and only if $Q_{ij} = 0$, where
$\bm{Z}_{(-ij)}$ is $\bm{Z}$ without the $i$th and $j$th observations.
In other words, the precision matrix structure completely determines 
the conditional dependence structure of pairs given others.
Since the undirected graph corresponding to a GMRF is retained in 
the resulting TGMRF, the equivalence of pairwise Markov property,
local Markov property, and global Markov property for a GMRF 
\citep{rue:held} are equivalent for a TGMRF.

In the sequel, a TGMRF with marginal distributions $\bm{F}$ and
precision matrix $\bm{Q}$ in the original GMRF scale is denoted
as $\mathrm{TGMRF}_n(\bm{F}, \bm{Q})$.
Matrix $\bm{Q}$ is not to be interpreted as precision but as 
a dependence matrix which characterizes the dependence structure.
This property can be exploited in modeling practice to construct
the precision matrix based on conditional dependences.

\section{Spatial Generalized Linear Mixed Models}
\label{s:glmm}

The TGRF and TGMRF open a new avenue of random field models such as
gamma field, beta field, and their Markov versions, which can be
incorporated into the GLMM framework for modeling spatial dependence.
Our departure point is the traditional GLMM with spatial random effects.

% \subsection{GLMM with TGRF}

Suppose that we observe $(Y_i, \bm{X}_i)$ at sites $i = 1, \ldots, n$,
where $Y_i$ is the response variable and $\bm{X}_i$ a $q \times 1$
vector of covariates that correspond to response $Y_i$ at site $i$.
Let $\bm{e} = (e_1, \ldots, e_n)^{\top}$ be a vector of unobserved random
effects with joint distribution $H$, which introduces spatial dependence.
A spatial GLMM assumes that, given ($\bm{X}_i, e_i$), $i = 1, \ldots, n$,
the observations $Y_i$'s are independent with a distribution from the
exponential family.
Let $\mu_i = E(Y_i|\bm{X},\bm{e})$, where
$\bm{X} =(\bm{X}_1,\ldots,\bm{X}_n)^{\top}$ is the matrix of covariates.
The conditional expectation $\mu_i$ is connected to the covariate
$\bm{X}_i$ and random effect $e_i$ through a fixed link function $g$:
\begin{equation}
\label{eq:GLMM}
   g(\mu_i) = \eta_i + e_i,
\end{equation}
where $\eta_i = \bm{X}_i^{\top}\beta$ is the fixed effect, and $\beta$ is
a $q\times 1$ vector of regression coefficients of covariates $\bm{X}_i$.
The dependence among random effects $\bm{e}$ determines the spatial
dependence among conditional means $\bm{\mu} = (\mu_1,\ldots,\mu_n)^{\top}$.
Therefore, to fully specify a spatial GLMM, it is necessary to specify
both the link function $g$ and the joint distribution $H$ of $\bm{e}$.
Commonly, $H$ is chosen to be a multivariate normal distribution
with mean zero and covariance matrix $\bm{\Sigma}$.

Instead of introducing dependence among $\bm{\mu}$ through
the joint distribution $H$ of random effects $\bm{e}$,
we propose to specify a random field directly for $\bm{\mu}$.
Specifically, our model for $\bm{\mu}$ is
\begin{equation}
  \label{eq:mu}
 \bm{\mu} \sim \mathrm{TGRF}_n(\bm{F},\bm{\Psi}),
\end{equation}
where $\bm{F} = (F_1,\ldots,F_n)$, $F_i$ is the marginal distribution
of $\mu_i$, and $\bm{\Psi}$ is the dispersion matrix characterizing
the dependence structure of the underlying Gaussian copula.
For independent data, in which case $\bm{\Psi}$ is the identity matrix,
this specification reduces to a class of GLMMs where the distribution
of conditional mean $\mu_i = g^{-1}(\eta_i+e_i)$, instead of
random effect $e_i$, is specified. % \citep{Prates:Dey:Yan:2011}.
Our specification here is more general in that it incorporates
dependence among all or part of $\mu_i$'s through Gaussian copulas.

The new model~\eqref{eq:mu} specifies the distribution of
$\bm{\mu}$ through marginal distributions $\bm{F}$ and
a Gaussian copula with dispersion matrix $\bm{\Psi}$.
It encompasses any model constructed from a link function
$g$ and $H = N_n(\bm{0}, \bm{\Sigma})$ as a special case 
where $F_i$ is the distribution function of 
$\mu_i = g^{-1}(\eta_i + e_i)$, $i = 1, \ldots, n$,
and $\bm{\Psi}$ is the correlation matrix of $\bm{\Sigma}$.

The TGRF model for $\bm{\mu}$ provides a natural choice
for the conditional means in hierarchical spatial models.
For instances, one can use gamma margins for Poisson intensities
and beta margins for Bernoulli rates, that can in turn be used,
respectively, to model spatial count data or spatial binary data.
The wide range of marginal distributions offer models that
cover the traditional models as special cases and many more
\citep[e.g.][]{Prates:Dey:Yan:2011}.
The spatial dependence is completely characterized by the
Gaussian copula, parameterized by the dispersion matrix $\bm{\Psi}$.
For geostatistical modeling, where the observations sites may be
irregularly spaced, one can parameterize $\bm{\Psi}$ using,
for examples, the exponential, spherical, or Mat\'ern structures
\citep[Ch.2]{Banerjee:Carlin:Gelfand:2004}.

% \subsection{GLMM with TGMRF}

Replacing the TGRF in model~\eqref{eq:mu} with a TGMRF,
we model the conditional means $\bm{\mu}$ by
\begin{equation}
  \label{eq:mu:tgmrf}
 \bm{\mu} \sim \mathrm{TGMRF}_n(\bm{F}, \bm{Q}),
\end{equation}
where the spatial dependence is characterized by $\bm{Q}$,
the precision matrix of the Gaussian copula.
Since the copula is invariant to scale changes, we do not
require that $\bm{Q}^{-1}$ is a correlation matrix as long
as $\bm{Q}$ is scale free, s.p.d. precision matrix.

Parameterization of $\bm{Q}$ is crucial and we propose
to used the structure of the precision matrix of a 
conditional autoregressive (CAR) model \citep{Besag:1974}.
In a CAR model, the precision matrix is defined as $\bm{Q} / \nu$,
where $\bm{Q}$ determines the structure and $\nu$ is a scale parameter.
The scale $\nu$ is not needed in our TGMRF model in~\eqref{eq:mu:tgmrf}.
The structure $\bm{Q}$ is defined in such way that $Q_{ij}$ is nonzero
if and only if site $i$ and site $j$ are neighbors of each other.
To assure symmetry and positive definiteness, $\bm{Q}$ is defined as
\begin{equation}
\label{eq:Q}
\bm{Q}=\bm{M}^{-1}(\bm{I} - \rho \bm{W}),
\end{equation}
where $\bm{M}^{-1}$ is a diagonal matrix whose $i$th diagonal elements
equal to $n_i$, the number of neighbors of site $i$, $\bm{I}$ is the
identity matrix, $\rho$ is a spatial dependence parameter, and $\bm{W}$
is a weight matrix providing contrasts of all neighbors to each site.
Weight matrix $\bm{W}$ is determined by the neighboring structure
and is of the form
\begin{equation*}
W_{ij} =
\begin{cases}
  1/n_i,  & i \sim j,\\
  0,       & \mbox{otherwise,}
\end{cases}
\end{equation*}
where $i \sim j$ indicates that site $i$ is a neighbor of site $j$.

% \subsection{Bayesian Inferences}

The proposed models fit naturally into the Bayesian framework.
With carefully chosen priors for the parameters, Markov chain
Monte Carlo (MCMC) algorithms can be developed to draw samples
from the posterior distribution of the parameters of interests
\citep[e.g.,][]{Gelm:Carl:Ster:Rubi:2003}.
To compare different models for the same data, we propose to
use the conditional predictive ordinate (CPO) criterion
\citep[e.g.,][]{Gelf:Dey:Chan:mode:1992,Dey:Chen:Chan:baye:1997}.
The summary statistic is the logarithm of the pseudo-marginal
likelihood (LPML), which is the summation of the log density of
leave-one-out marginal posterior distribution.
The performance of the CPO criterion in selecting the right
models will be studied through simulations.
The deviance information criterion (DIC) \citep{Spiegelhalter:etal:2002}
is an alternative Bayesian model selection criterion.
In our simulation studies, however, DIC had much higher variation 
than LPML and was outperformed in selecting the correct models. 
This might be explained by the fact that the DIC measures are highly 
dependent on the marginalization of the random effects, and become 
unstable when the distributions are nonnormal.

\section{Spatial Poisson Application}
\label{s:po}

Consider count data observed at $n$ sites in a spatial domain.
Let $Y_i$ be the count at site $i$, and with a $q\times 1$
covariate vector $\bm{X}_i$, $i = 1, \ldots, n$.
Poisson models are widely used for count data and the
Poisson intensities are often modeled by gamma distributions.
Few choices of gamma fields are available in the literature.
An exception is \citet{Wolp:Ichs:pois:1998}, where a doubly
stochastic process is used to construct positively autocorrelated
intensity measures for spatial Poisson point processes which
in turn are used to model the spatial count data.
The TGMRF models provides new gamma Markov random fields
to account for spatial dependence.

\subsection{TGMRF Models}
A GLMM introduces spatial dependence through a spatial random effect.
Conditioning on $\bm{\mu} = (\mu_1, \ldots, \mu_n)^\top$, the observed
spatial count data $Y_i$'s are assumed to be independent, and
each $Y_i$ is Poisson with mean $\mu_i$, $i = 1, \ldots, n$.
The most commonly used GLMM for spatial count data uses the
canonical log link on the Poisson intensities:
\begin{equation}
\label{eq:mu:lnorm}
\log \mu_i = \bm{X}_i^{\top} \beta + e_i,
\end{equation}
where $\beta$ is a $q \times 1$ regression coefficient vector,
$\bm{e} = (e_1, \ldots, e_n)^{\top}$ follows a GMRF with mean
zero and a s.p.d. precision matrix $\bm{\Omega}/\nu$, and $\nu > 0$
is a parameter controlling the scale of the variance.
Let $\sigma_i^2$ be the $i$th diagonal element of $\bm{\Omega}^{-1}$.
Let $\bm{F} = (F_1, \ldots, F_n)^{\top}$, where $F_i$ is
the distribution function of
\begin{equation}
  \label{eq:lnorm}
  \mathrm{LN}(\bm{X}_i^{\top} \beta, \quad \nu \sigma_i^2),
  \quad \nu > 0,
  \quad i = 1, \ldots, n,
\end{equation}
where $\mathrm{LN}(a, b)$ denotes a log-normal distribution
with mean $a$ and variance $b$ on the log scale.
It is clear that model~\eqref{eq:mu:lnorm} is a special case of
model~\eqref{eq:mu:tgmrf} with $\bm{Q} = V^{1/2} \bm{\Omega} V^{1/2}$ 
and $V = \diag(\sigma_1^2, \ldots, \sigma_n^2)$.

The TGMRF framework provides a new way to construct models for
$\bm{\mu}$ that incorporate spatial dependence and covariates.
The Gaussian copula of TGMRFs captures the spatial dependence.
Any positive continuous distribution can be used to specify the
marginal distribution of $\bm{\mu}$, and covariate effects
can be accommodated into its parameters. 
Changing $\bm{F}$ in model~\eqref{eq:mu:tgmrf} from log-normal to
other distribution functions with positive support leads to new models.
Gamma distribution is a natural choice for the margins.
Let $\Gamma(a, b)$ represent a gamma distribution with
shape parameter $a$ and scale parameter $b$, hence mean $ab$.
Covariates can be incorporated into either one of the two
parameters, resulting in two different gamma models as
long as there is at least one covariate.
The gamma scale model, hereafter the GSC model, incorporates
covariates into the scale parameter and defines the marginal
distribution $F_i$ as
\begin{equation}
  \label{eq:gamma:scale}
  \Gamma\left( 1/ \nu, \quad \nu \exp(\bm{X}_i^{\top} \beta) \right),
  \quad \nu > 0,
  \quad i = 1, \ldots, n.
\end{equation}
The gamma shape model, hereafter the GSH model, incorporates
covariates into the shape parameter and defines the marginal
distribution $F_i$ as
\begin{equation}
  \label{eq:gamma:shape}
  \Gamma\left(\exp(\bm{X}_i^{\top} \beta) / \nu, \quad \nu \right),
  \quad \nu > 0,
  \quad i = 1, \ldots, n.
\end{equation}
Under both models, the expectation of $\mu_i$ is the same,
$\exp(\bm{X}_i^{\top} \beta)$, but the parameter $\nu$ has
different interpretations and should not be compared directly.
TGMRF models with other marginal distribution for $\mu_i$s
can be constructed similarly.

There is a subtle difference between the log-normal 
model~\eqref{eq:lnorm}, hereafter the LN model, and the two 
gamma models~\eqref{eq:gamma:scale} and~\eqref{eq:gamma:shape}.
Unlike the gamma models, where the dependence structure 
does not interfere with the marginal models, the dependence 
structure $\bm{Q}$ enters the marginal distributions of 
$\mu_i$s through $\sigma_i^2$ in the LN model.
This implies that a different model could be constructed with
$F_i$ being the distribution of
\begin{equation}
  \label{eq:lnorm2}
  \mathrm{LN}(\bm{X}_i^{\top} \beta, \quad \nu),
  \quad \nu > 0,
  \quad i = 1, \ldots, n.
\end{equation}
The variance parameter $\nu$ could even incorporate covariates.
These model could be used as alternatives to the commonly
used LN model~\eqref{eq:lnorm} in the TGMRF framework.

\subsection{Abundance of \emph{Nenia tridens}}
\label{s:po:nt}

Because of their abundance and critical roles in nutrient
cycling, gastropods are of considerable ecological
importance in terrestrial ecosystems \citep{Mason:1970}.
In the Luquillo Mountains of Puerto Rico, \emph{Nenia tridens} is one of
the most abundant and widely distributed terrestrial gastropods in tabonuco
forest \citep{Willig:etal:1998:b,Bloch:Willig:2006,Willig:etal:2011}.
Indeed, the forest ecosystems of the Luquillo Mountains have a long history
of environmental study \citep[e.g.,][]{Brow:etal:1983,Reagan:Waide:1996},
resulting in deep understanding of the spatial and temporal dynamics
of populations, communities, and biogeochemical processes, especially as
they relate to natural and human disturbances \citep{Brokaw:etal:2011}.

\begin{figure}
  \centering
  \renewcommand{\subfigbottomskip}{0pt}\renewcommand{\subfigtopskip}{0pt}\renewcommand{\subfigcapskip}{0pt}\renewcommand{\subfigcapmargin}{0pt}
  \subfigure[]{\label{fig:lat}\includegraphics[width=0.3\textwidth]{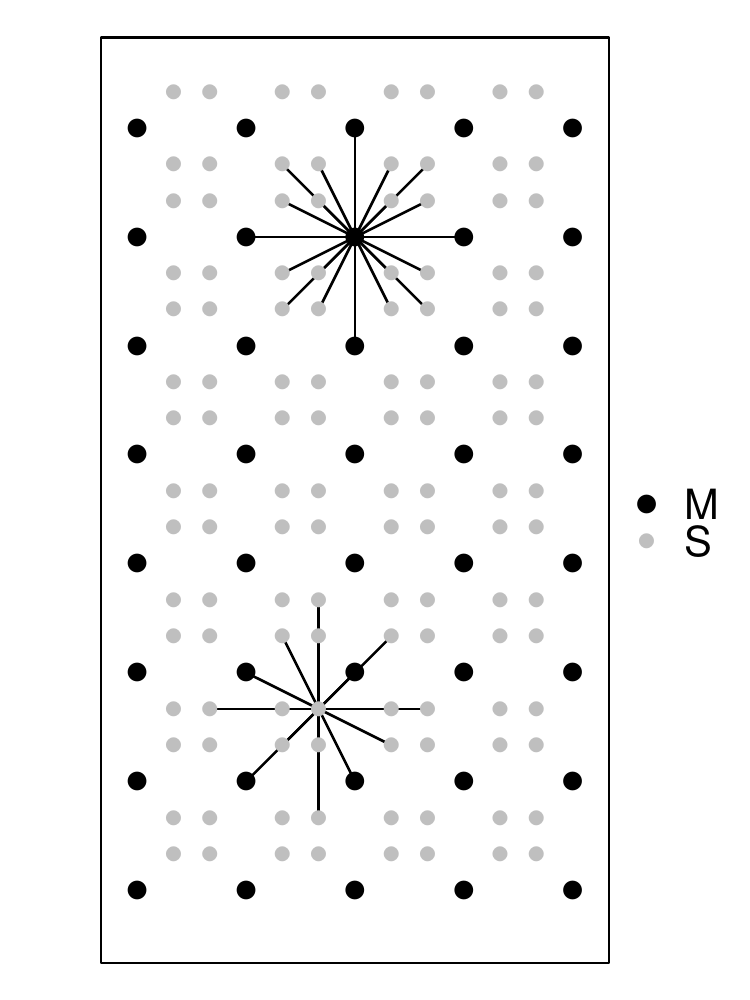}}
  \subfigure[]{\label{fig:nt}\includegraphics[width=0.3\textwidth]{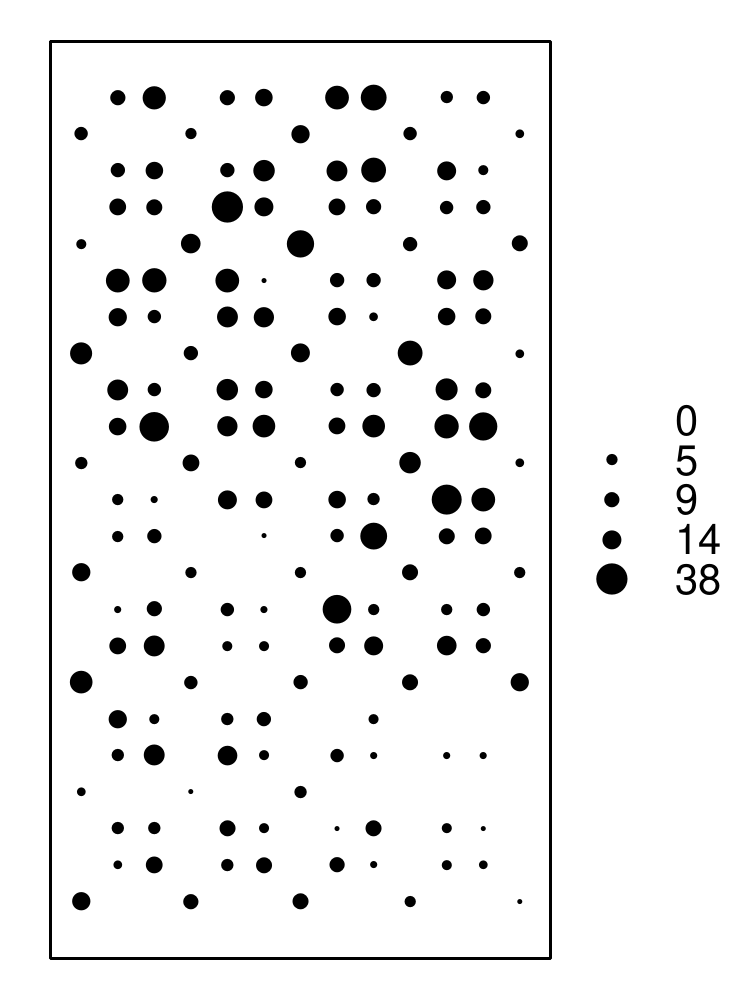}}
  \subfigure[]{\label{fig:gn}\includegraphics[width=0.3\textwidth]{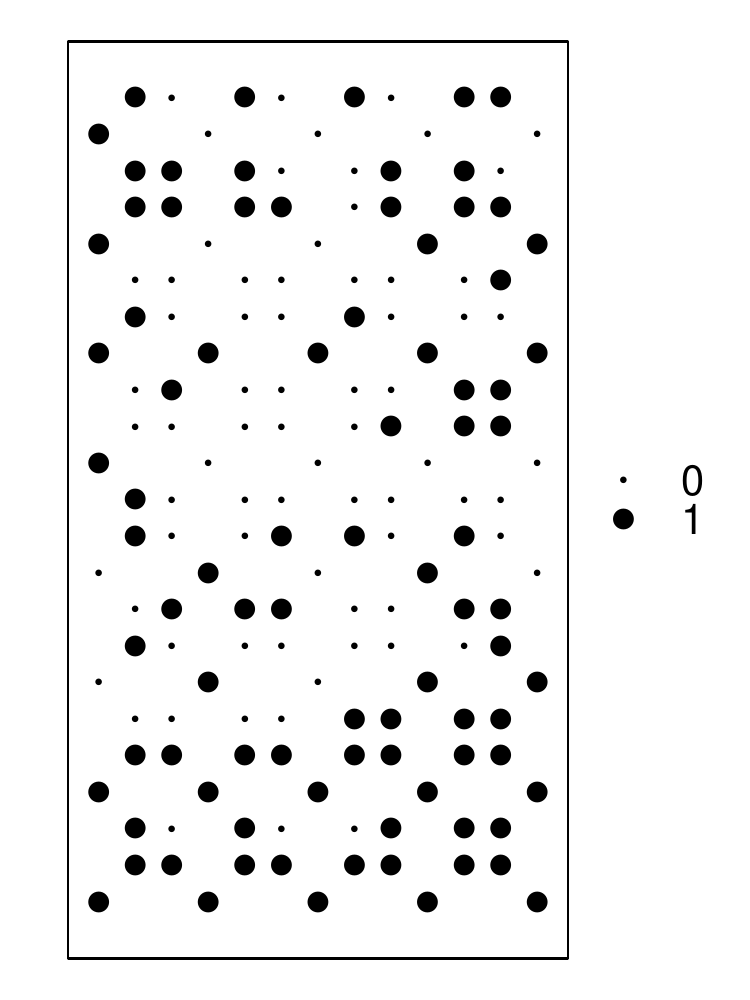}}
  \caption{(a) Lattice of sampling sites at the LFDP in 1995 and the neighbor structure for an internal major site and an internal supplementary site, labeled by M and S, respectively. (b) Abundance of \emph{N. tridens}. (c) Presence/absence of \emph{G. nigrolineata}.}
  \label{fig:appl}
\end{figure}

Abundance data of \emph{N. tridens} were collected from the Luquillo
Forest Dynamics Plot (LFDP), a 16 hectare grid in tabonuco forest
($18^{\circ} 20'$N and $65^{\circ} 49'$W), during the wet season of
1995 at each of 160 circular sites (3 m radius) on an lattice.
As shown in Figure~\ref{fig:lat}, there are 40 major sites in dark,
60 meters apart, and 120 supplementary sites in gray, 20 meters
apart, placed inside the squares formed by the 40 major sites.
Therefore, the data is available on a regular but sparse lattice.
To define the graph for the TGMRF model, any two sites within
60 meters are considered neighbors, which results in different
number of neighbors for major sites and for supplementary sites.
Also shown in Figure~\ref{fig:lat} are an internal major site
connected to its 20 neighbors and an internal supplementary site
connected to its 16 neighbors.

The abundance of \emph{N. tridens} at each site was the minimum number
known alive from four nocturnal surveys based on well established
protocols on the LFDP \citep{Willig:etal:1998:b,Bloch:Willig:2006}.
The observed count over the lattice is displayed in Figure~\ref{fig:nt}.
Possible covariates were topographic and habitat characteristics at each site.
There were two topographic variables, elevation and slope.
Four habitat variables were quantity of litter, canopy openness,
apparency of sierra palm, and plant apparency.
Quantity of litter was the mean number of leaves on the forest floor
from each of four locations that were sampled at each site along
mid-points of the radii from the center of the circle, arranged along
cardinal compass directions (cardinal points).
Canopy openness was the amount of light that penetrates to the understory
(1.5 m above the forest floor) based on the mean number of open cross-hairs
on a gridded densiometer, quantified from the four cardinal points.
Plant apparency measured the volume of space in the understory that
was occupied by plants using a plant apparency device at each of the
four cardinal points, which captured the number of foliar
intercepts along each of two perpendicular 1.0 m dowels placed at
0.5 m intervals from ground level to 3 m of height.
Apparency of sierra palm measured specifically the apparency of
\emph{Prestoea acuminata}, a preferred substrate and food of
\emph{N. tridens}.

We fitted Poisson regressions to the abundance data of
\emph{N. tridens} with three TGMRF models: LN, GSC, and GSH.
For each model, the precision matrix $Q$ of the Gaussian 
copula was specified with~\eqref{eq:Q} from the CAR model.
The prior distributions of regression coefficients $\beta_i$,
$i = 1, \ldots, q$, are independent $N(0, 1/\tau)$, with $\tau$ = 0.01.
The prior distribution of the scale or shape parameter
$\nu$ is specified as $\Gamma(\kappa_1, \kappa_2)$,
with $\kappa_1 = 0.01$ and $\kappa_2 = 100$.
These priors are set to be proper but vague to allow
the posterior estimates to be mainly data driven.
Because we expected positive spatial dependence, a $U(0,1)$ prior
is put on the spatial dependence parameter, $\rho$ for the CAR model.

The GSH model had the largest LPML, $-$482.12, followed by the 
GSC model ($-$483.90) and the LN model ($-$491.15).
These results suggest that, The GSH model and the GSC model
performed fairly closely, with the former being mildly preferred.
The GSH model provides considerably better fit than the
traditional LN model with a 9.1 difference in LPML.
Since both models have the same number of parameters,
the log Bayes factor is approximately twice LPML
difference asymptotically \citep{Gelf:Dey:baye:1994}.
From the rules suggested by \citet{Kass:Raft:baye:1995}, a log Bayes
factor of 18.2, which falls in the category of 10 or higher, provides
``very strong'' evidence in favor of the GSH model over the LN model.
As will be seen in our simulation study, when the true model was the
GSH model, 38 out of 100 replicates had LPML differences of greater
than 9.1 between the fitted GSH model and the fitted LN model;
when the true model was the either one of the two LN models considered,
however, this rate became 0 out of 100.

\begin{table}
\caption{Posterior point estimates and 95\% HPD credible intervals of the parameters in the Poisson regression for the abundance of \emph{N. tridens} with the GSH model and the traditional LN model. The regression coefficients are in the order of intercept, elevation, slope, quantity of litter, canopy openness, plant apparency, and apparency of sierra palm.}
\label{t:nenia}
\centering
  \begin{tabular}{c cc cc}
    \toprule
    Parameters & \multicolumn{4}{c}{Specified Model}\\
    \cmidrule(lr){2-5}
     & \multicolumn{2}{c}{GSH} & \multicolumn{2}{c}{LN}\\
    \cmidrule(lr){2-3}\cmidrule(lr){4-5}
     & Estimate & 95\% HPD & Estimate & 95\% HPD \\
    \midrule
    \multicolumn{5}{l}{Regression coefficients}\\
    $\beta_0$  & 1.990    & (0.841, 2.731)       & 2.137       & (1.434, 2.839) \\
    $\beta_1$  & $-$0.062 & ($-$0.293, 0.174)    & $-$0.037    & ($-$0.329, 0.226) \\
    $\beta_2$  & $-$0.046 & ($-$0.179, 0.081)    & $-$0.045    & ($-$0.163, 0.064) \\
    $\beta_3$  & $-$0.103 & ($-$0.238, 0.020)    & $-$0.105    & ($-$0.232, 0.015)\\
    $\beta_4$  & $-$0.143 & ($-$0.292, $-$0.014) & $-$0.129    & ($-$0.252, 0.007) \\
    $\beta_5$  & 0.027    & ($-$0.100, 0.161)    & 0.021       & ($-$0.107, 0.139) \\
    $\beta_6$  & 0.033    & ($-$0.104, 0.161)    & 0.024       & ($-$0.105, 0.149) \\
    \multicolumn{5}{l}{Scale and spatial dependence parameters}\\
    $\nu$      & 4.924    & (3.902, 7.312)       & 5.134       & (3.534, 6.861) \\
    $\lambda$  & 0.951	  & (0.851, 0.996)       & 0.953       & (0.851, 0.999) \\
  \bottomrule
\end{tabular}
\end{table}

The posterior point estimates and 95\% highest posterior density
(HPD) credible intervals of the parameters from the GSH model and 
the traditional LN model are summarized in Table~\ref{t:nenia}.
The two models lead to qualitatively the same conclusions.
Neither elevation nor slope was found to have a
significant effect on the abundance of \emph{N. tridens}.
Of the habitat variables, only canopy openness is negatively significant.
More openness in the canopy implies fewer trees and dryer soil, which
are not the preferred habitat condition by the \emph{N. tridens}.
The marginal scale parameter $\nu$ is estimated to be 4.924.
The spatial dependence parameter $\rho$ is estimated as 0.951,
with a HPD interval away from zero, which indicates
a higher spatial dependence in the model is needed.

\subsection{Simulation Study}
\label{s:po:ss}

To assess the fitting capacity of the TGMRF models, the properties
of the Bayesian inferences, and the effectiveness of LPML as a model
comparison criterion in this context, we conducted a simulation study
using the lattice and neighbor structure in Figure~\ref{fig:lat}.
Each of the three models was used as data generating models.
In addition to the intercept, one covariate was generated from $N(0, 1)$,
and the true covariate coefficient vector was $\beta = (1.0, 0.7)$.
The precision matrix of the TGMRF took the form of~\eqref{eq:Q}
for the CAR model, with $\rho = 0.8$.
The parameter $\nu$, which is related to the variance in all models,
was set at $\nu = 2$, although it has completely different meanings.
With $\nu = 2$, the gamma scale model and the gamma shape model
appeared to be more similar to each other than to the log-normal model.
To make a more interesting comparison, a second log-normal model
was also used to generate data, where $\nu = 6.5$ was chosen because
it provides good approximation to the gamma scale model with $\nu = 2$.
In summary, we had a total of four data generating models:
two LN models LN1 and LN2, one GSC model, and one GSH model.

For each data generating model, we generated 100 datasets,
and fit each dataset with all three proposed TGMRF models.
In each fitting process, a vague prior, $\Gamma(0.01, 100)$,
was set for the dispersion parameter $\nu$, and an uninformative
$U(0, 1)$ prior was set for the spatial dependence parameter $\rho$.
Independent $N(0, 100)$ priors were set on regression coefficients $\beta$.
Table~\ref{tab:pois:sim} summarizes the mean and standard deviations of the
Bayesian estimate of the parameters and LPML from the 100 replicates.

\begin{table}
\caption{Summaries of posterior mean, standard deviations (SD), and LPML from 100 replicates in the simulation of spatial Poisson regression.}
\label{tab:pois:sim}
\centering
\begin{tabular}{c cc rrrrrr}
  \toprule
  True  & Param & True & \multicolumn{6}{c}{Specified Model}\\
  \cmidrule(lr){4-9}
  Model &            & Value & \multicolumn{2}{c}{LN} & \multicolumn{2}{c}{GSC} & \multicolumn{2}{c}{GSH} \\
  \cmidrule(lr){4-5}\cmidrule(lr){6-7}\cmidrule(lr){8-9}
  & & & Mean & SD & Mean & SD & Mean & SD \\
  \midrule
  LN1 & $\beta_0$ & 1.00 & 0.99 & 0.10 & 1.08 & 0.09 & 1.11 & 0.14 \\
  & $\beta_1$ & 0.70 & 0.70 & 0.06 & 0.70 & 0.05 & 0.67 & 0.06 \\
  & $\rho$ & 0.80 & 0.53 & 0.26 & 0.55 & 0.26 & 0.55 & 0.26 \\
  & $\nu$ & 2.00 & 2.32 & 0.73 & 6.35 & 1.53 & 0.84 & 0.46 \\
  \vspace{3mm}
  & LPML &  & $-$331.90 & 10.76 & $-$332.55 & 10.87 & $-$335.57 & 11.11 \\
  LN2 & $\beta_0$ & 1.00 & 0.98 & 0.16 & 1.33 & 0.18 & 1.46 & 0.23 \\
  & $\beta_1$ & 0.70 & 0.70 & 0.08 & 0.70 & 0.07 & 0.58 & 0.08 \\
  & $\rho$ & 0.80 & 0.60 & 0.22 & 0.61 & 0.21 & 0.64 & 0.22 \\
  & $\nu$ & 6.50 & 6.97 & 1.42 & 2.00 & 0.41 & 3.50 & 1.16 \\
  \vspace{3mm}
  & LPML &  & $-$362.90 & 15.06 & $-$365.69 & 14.78 & $-$371.35 & 15.77 \\
  GSC & $\beta_0$ & 1.00 & 0.76 & 0.18 & 0.99 & 0.13 & 1.09 & 0.19 \\
  & $\beta_1$ & 0.70 & 0.70 & 0.08 & 0.70 & 0.07 & 0.61 & 0.07 \\
  & $\rho$ & 0.80 & 0.59 & 0.23 & 0.59 & 0.23 & 0.60 & 0.23 \\
  & $\nu$ &  2.00 & 6.34 & 1.30 & 2.24 & 0.53 & 2.18 & 0.73 \\
  \vspace{3mm}
  & LPML &  & $-$336.34 & 14.96 & $-$335.38 & 14.47 & $-$340.95 & 15.12  \\
  GSH & $\beta_0$ & 1.00 & 0.71 & 0.20 & 1.00 & 0.14 & 0.99 & 0.18 \\
  & $\beta_1$ & 0.70 & 0.77 & 0.08 & 0.71 & 0.08 & 0.70 & 0.07  \\
  & $\rho$ & 0.80 & 0.64 & 0.21 & 0.62 & 0.21 & 0.63 & 0.21 \\
  & $\nu$ &  2.00 & 7.09 & 1.31 & 1.95 & 0.48 & 2.17 & 0.60 \\
  & LPML &  & $-$335.86 & 17.27 & $-$335.96 & 16.60 & $-$328.81 & 17.35  \\
  \bottomrule
\end{tabular}
\end{table}

When the model was correctly specified, the true values
of the regression coefficients were recovered very well.
The estimates seems to be upward biased for the dispersion parameter $\nu$
but downward biased for the dependence parameter $\rho$, suggesting that
spatial dependence and spatial heterogeneity are hard to identify.
When the model was misspecified, the regression coefficient estimates
were still recovered reasonably well, especially in the GSC model
and the GSH model, probably because the mean of $\bm{\mu}$
was still correctly specified, regardless of the misspecified model.
In all cases, the average of the LPML statistic was higher for correctly
specified models than for the misspecified models, with similar variation
under different models.

\begin{table}
\caption{Frequencies of selected model using the LPML statistics for the 100 simulated datasets.}
\label{t:cpo}
\centering
\begin{tabular}{cccc}
   \toprule
  True model & \multicolumn{3}{c}{Frequency selected}\\
  \cmidrule(lr){2-4}
  & LN & GSC & GSH \\
  \midrule
  LN1 & 59 & 29 & 12 \\
  LN2 & 77 & 16 & 7 \\
  GSC & 34 & 59 & 7 \\
  GSH & 6 & 5 & 89 \\
  \bottomrule
\end{tabular}
\end{table}

To gain a clearer picture on model comparison using LPML, we summarize
the frequencies of the models selected with the highest LPML from all
100 replicates under each of the four models (Table~\ref{t:cpo}).
The criterion seems to be very effective when the true model was
the GSH model, correctly selecting the true model 89 times.
When the true model was LN1 or GSC, the correct model was selected 59 
times in either case with our sample size, while the alternative
GSC model or LN model was selected 29 and 34 times, respectively;
the GSH model was selected only 12 and 7 times, respectively.
This indicates that the LN model and the GSC model provides good 
approximation to each other, similar to their well known similarity
in univariate modeling without covariates and spatial concerns;
a large sample would be necessary to distinguish them effectively.
With our sample size, when the true model was LN2, the LPML was 
able to differentiate the LN model better from the GSC
model, correctly selecting the LN model 77 times.
Therefore, the similarity between the GSC model and the
LN model appear to be different under different scenarios.
The GSH model seems to have specific characteristics that make it
further away from both the LN model and the GSC model in the model space.

\begin{figure}
  \centering
  \renewcommand{\subfigbottomskip}{0pt}\renewcommand{\subfigtopskip}{0pt}\renewcommand{\subfigcapskip}{0pt}\renewcommand{\subfigcapmargin}{0pt}
  \subfigure[]{\label{fig:pois}\includegraphics[width=0.48\textwidth]{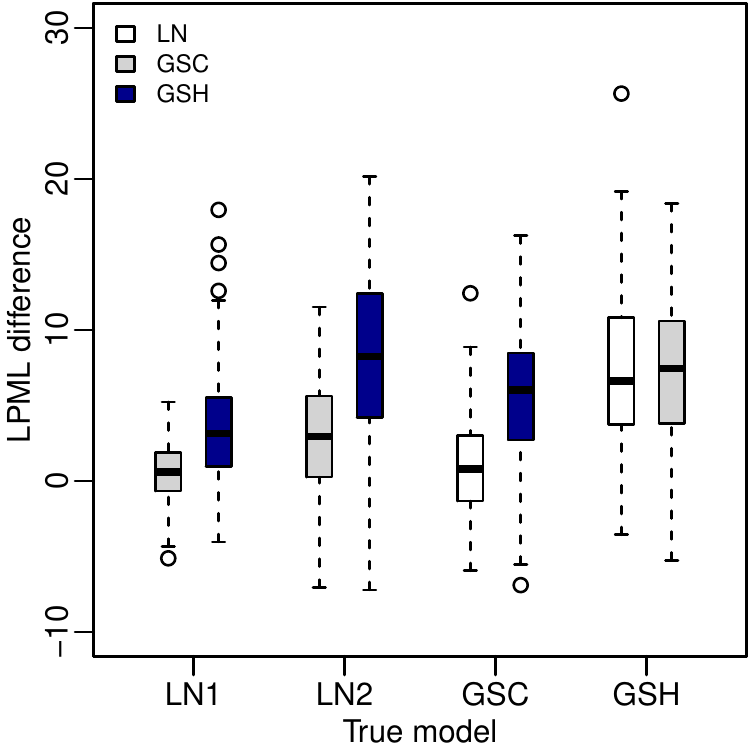}}
  \subfigure[]{\label{fig:bern}\includegraphics[width=0.48\textwidth]{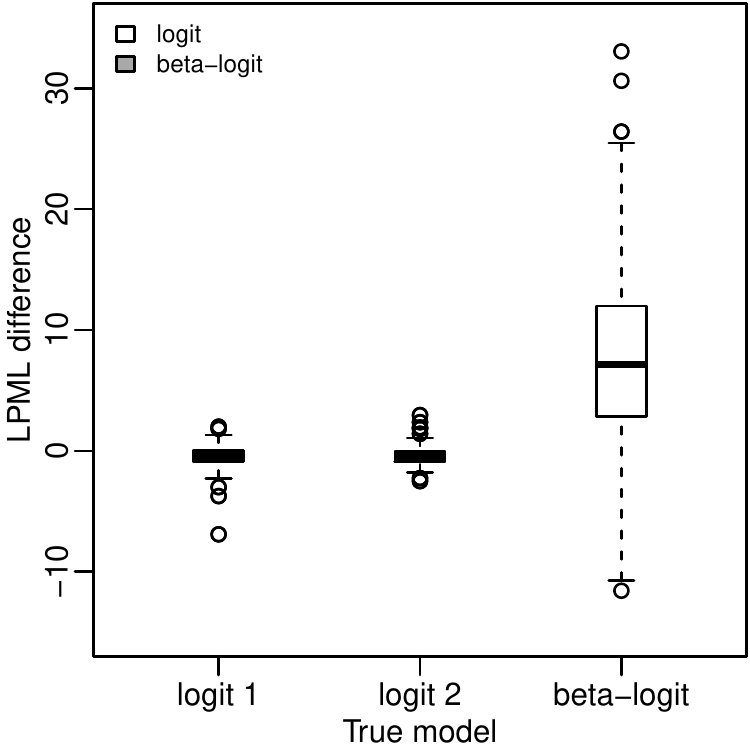}}
  \caption{LPML difference between the correct model and misspecified models. (a) Poisson simulation. (b) Bernoulli simulation.}
  \label{fig:lpml}
\end{figure}

A closer look at the difference in LPML across models is through box plots.
Figure~\ref{fig:pois} presents the box plots of the difference in LPML
between the correct model and two misspecified models for each true model.
The magnitude of the differences provides guidance in practice on what
models are similar to each other and on how big a difference is important.
In the spatial setting we considered, the LN model and the GSC model
were very similar, as seen from the boxes centered near zero.
The majority of each box plot is well above $-5$, suggesting that if
the LPML of one model is observed to be higher than that of another model
by 5, then it is very unlikely that the other model is the true model.

\section{Spatial Bernoulli Application}
\label{s:bin}
Consider presence/absence data at $n$ sites in a spatial domain.
Let $Y_i$ be $1$ if presence is observed and $0$ otherwise
at site $i$, with a $q\times 1$
covariate vector $\bm{X}_i$, $i = 1, \ldots, n$.

\subsection{TGMRF Models}
Conditioning on $\bm{\mu} = (\mu_1, \ldots, \mu_n)^\top$,
the observed data $Y_i$'s are assumed to be independent, and
each $Y_i$ is Bernoulli with mean $\mu_i$, $i = 1, \ldots, n$.
The traditional spatial GLMM for binary data is
\begin{equation}
\label{eq:mu:logit}
\mathrm{logit}(\mu_i) = \bm{X}_i^{\top} \beta + e_i,
\end{equation}
where $\beta$ is a $q\times 1$ regression coefficient vector,
$\bm{e} = (e_1, \ldots, e_n)^{\top}$ follows a GMRF with mean
zero and precision matrix $\bm{\Omega}/\nu$, and $\nu > 0$
is a parameter controlling the scale of the variance.
Let $\sigma_i^2$ be the $i$th diagonal element of $\bm{Q}^{-1}$.
Let $\bm{F} = (F_1, \ldots, F_n)^{\top}$, where $F_i$ is
the distribution function
$\mu_i = \mathrm{logit}^{-1}(\bm{X}_i^{\top} \beta + e_i)$
$i = 1, \ldots, n$.
Then, model~\eqref{eq:mu:logit} is a special case of
model~\eqref{eq:mu:tgmrf} with $\bm{Q} = V^{1/2} \bm{\Omega} V^{1/2}$
and $V = \diag(\sigma_1^2, \ldots, \sigma_n^2)$.

Changing $\bm{F}$ in model~\eqref{eq:mu:tgmrf} to any distribution
function defined over the $(0,1)$ support leads to new models.
Covariate effects can be accommodated into the marginal parameters.
Spatial dependence is modeled through the Gaussian copula
with dispersion matrix $\bm{\Omega}^{-1}$.

The beta distribution is a natural choice for the margins.
Let Beta$(\nu p, \nu (1-p))$ represent a beta distribution
with mean parameter $p$ and dispersion parameter $\nu$.
Covariates can be incorporated into the mean parameter $p$ using any
transformation function from $\Re$ to (0,1) \citep[e.g.,][]{Ferrari:Neto:2004}.
We propose a beta-logit model that incorporates covariates
into the mean parameter $p$ using a inverse logit transformation
and defines marginal distribution $F_i$ as
\begin{equation}
  \label{eq:beta:logit}
  \mathrm{Beta}\left[\nu \frac{\exp(\bm{X}_i^{\top} \beta)}{\exp(\bm{X}_i^{\top} \beta)+1}, \quad \nu \left\{1 -  \frac{\exp(\bm{X}_i^{\top} \beta)}{\exp(\bm{X}_i^{\top} \beta)+1}\right\} \right],
  \quad \nu > 0,
  \quad i = 1, \ldots, n.
\end{equation}

There is again a subtle difference between the logit model~\eqref{eq:mu:logit}
in comparison with the beta-logit model~\eqref{eq:beta:logit}.
In the logit model~\eqref{eq:mu:logit}, parameters in the dependence
structure $\bm{Q}$ enters the marginal distribution, whereas it does not
do so in the beta model.

\subsection{Presence of \emph{Gaeotis nigrolineata}}
\label{s:bin:ga}

\emph{Gaeotis nigrolineata} is a common terrestrial gastropod in
tabonuco forest of the Luquillo Mountains of Puerto Rico
\citep{Willig:etal:1998:b,Bloch:Willig:2006,Willig:etal:2011}.
Its spatial distributions is believed to be associated with
the abundance of live sierra palms, its preferred substrate.
Generally, \emph{Gaeotis nigrolineata} is less abundant than is \emph{N. tridens}.
It often occurs in low numbers, and is characteristically absent
from a significant proportion of the sites across the LFDP.
Therefore, it is more suitable to analyze the presence/absence
data for this taxon.

The presence/absence data were obtained by dichotomizing the
abundance of \emph{G. nigrolineata}, which were determined in the same
manner as described for \emph{N. tridens} (Section~\ref{s:po:nt}).
In particular, we have one for presence and zero for absence at each site.
The distribution of incidences for \emph{G. nigrolineata}
is apparently heterogeneous with spatial clustering
across the data collection lattice; see Figure~\ref{fig:gn}.
All but one of the covariates as described in Section~\ref{s:po:nt}
were used to model spatial dynamics of \emph{G. nigrolineata}.
Since \emph{G. nigrolineata} does not live or feed in the leaf litter,
quantity of litter was not included as a covariate in its analysis.

We fitted Bernoulli regressions for presence/absence data of
\emph{G. nigrolineata} with two TGMRF models: logit and 
beta-logit with precision matrix of the CAR model.
Prior distributions for the models parameters were selected
the same as those described in Section~\ref{s:po:nt}.

The LPML values were $-$99.31 and $-$103.51 for the 
beta-logit model and the logit model, respectively.
Therefore, using a CAR dependence structure, the beta-logit 
model fits better than the traditional logit model.
The approximate log Bayes factor was 8.4, which falls in the 
category of [6, 10) suggested by \citet{Kass:Raft:baye:1995},
``strong'' evidence favoring the beta-logit model over the logit model.
To be seen in our simulation study, when the true model was beta-logit,
68 out of 100 replicates had LPML differences of greater than 4.2
between the fitted beta-logit model and the fitted logit model, but
the rate was 1 or 0 out 100 when the true model was a logit model.

The posterior point estimates and 95\% HPD credible intervals
for parameters in both models are summarized in Table~\ref{t:gaeotis}.
The conclusions of the two models are virtually the same.
Neither elevation nor slope had a significant effect on the incidence
of \emph{G. nigrolineata}, as in the case for \emph{N. tridens}.
Of the habitat characteristics, only plant apparency had a significantly
negative effect on the incidence of \emph{G. nigrolineata},
That is, the greater the volume of vegetation in the understory
of the forest, the lower the abundance of \emph{G. nigrolineata}.
The apparency of sierra palm, which measures the preferred substrate
for the \emph{G. nigrolineata}, was found to be almost positively
significant with the 95\% HPD credible interval barely including zero.
The negative effect of plant apparency was surprising but the paradox may be
resolved if high plant apparency in the understory indicates the presence
of an opening in the canopy, and attendant temperatures (high) and
humidities (low) outside of the fundamental niche of \emph{G. nigrolineata},
precluding its presence even though its preferred substrate may be common.
The spatial dependence parameter $\rho$ is estimated as 0.760 and
0.803 in the two models, respectively, indicating strong spatial
dependence within neighbors areas.

\begin{table}
\caption{Posterior point estimates and 95\% HPD credible intervals of the parameters in the Bernoulli regressions for the presence/absence of \emph{G. nigrolineata} with the beta-logit model and the traditional logit model using the CAR dependence structure. The regression coefficients are in the order of intercept, elevation, slope, canopy openness, plant apparency, and apparency of sierra palm.}
\label{t:gaeotis}
  \centering
  \begin{tabular}{c cc cc}
    \toprule
    Parameters & \multicolumn{4}{c}{Specified Model}\\
    \cmidrule(lr){2-5}
     & \multicolumn{2}{c}{beta logit} & \multicolumn{2}{c}{logit}\\
    \cmidrule(lr){2-3}\cmidrule(lr){4-5}
     & Estimate & 95\% HPD & Estimate & 95\% HPD \\
    \midrule
    \multicolumn{5}{l}{Regression coefficients}\\
    $\beta_0$  & 0.298    & ($-$0.481, 1.253)    & 0.226    & ($-$1.733, 1.839) \\
    $\beta_1$  & 0.326    & ($-$0.174, 0.777)    & 0.527    & ($-$0.578, 1.744) \\
    $\beta_2$  & 0.087    & ($-$0.238, 0.428)    & 0.134    & ($-$0.521, 0.803) \\
    $\beta_3$  & $-$0.014 & ($-$0.340, 0.344)    & $-$0.051 & ($-$0.715, 0.704) \\
    $\beta_4$  & $-$0.500 & ($-$0.887, $-$0.137) & $-$0.894 & ($-$1.784, $-$0.174) \\
    $\beta_5$  & 0.270    & ($-$0.074, 0.644)    & 0.538    & ($-$0.200, 1.330) \\
    \multicolumn{5}{l}{Scale and spatial dependence parameter}\\
    $\nu$      & 1.699    & (0.235, 4.551)       & 92.808   & (0.200, 234.300) \\
    $\rho$     & 0.760    & (0.264, 0.998)       & 0.803    & (0.326, 0.999) \\
  \bottomrule
\end{tabular}
\end{table}

It is worth noting that although the beta-logit model agrees with the
logit model in the directions of the covariates effects, it has much
smaller widths in the HPD credible interval does the logit model.
This indicates better precision of the estimating the coefficients.
The $\nu$ parameter does not have the same interpretation in
the two models, and, hence, they are not directly comparable.
Nevertheless, from Table~\ref{t:gaeotis}, we can see that for
the beta-logit model where $\nu$ is a marginal overdispersion
parameter, $\nu$ is more identifiable with a small HPD interval.
For the logit model, the $\nu$ parameter is the marginal variance.
The estimate implies a standard deviation of 9.634 with a wide
95\% credible interval of (0.447, 15.297).
On the log scale of $\mu$, such a magnitude of variation may
not mean much on the original scale of $\mu$ since the log
transformation explodes at zero, and this may explain
the poor identification of the spatial logit model.

\subsection{Simulation Study}
\label{s:bin:ss}

A simulation study was conducted for the spatial Bernoulli regressions.
Both the logit model and the beta-logit model with the CAR
dependence structure were used to generate data.
Except for the response variable, the simulation setup was
the same as that in Section~\ref{s:po:ss} with model parameters
$\beta = (1.0, 0.7)$, $\rho = 0.8$ and $\nu = 2$.
Again, since $\nu$ has different interpretation in the two models,
a second logit model with $\nu = 1$ was also used to generate data
in attempt to approximate the beta-logit model with $\nu = 2$.
For each of three true models, we generated 100 datasets,
and fit each dataset with each of two TGMRF models.
The priors were chosen in the same manner as Section~\ref{s:po:ss}.
Table~\ref{tab:bin:sim} summarizes the posterior mean and standard
deviations estimates from 100 replicates.

\begin{table}
\caption{Summaries of posterior mean, standard deviations (SD) and LPML from 100 replicates in the simulation of spatial Bernoulli regression.}
\label{tab:bin:sim}
\centering
\begin{tabular}{c cr rrrr}
  \toprule
  True  & Param & True & \multicolumn{4}{c}{Specified Model}\\
  \cmidrule(lr){4-7}
  Model &            & Value & \multicolumn{2}{c}{logit} & \multicolumn{2}{c}{beta-logit} \\
  \cmidrule(lr){4-5}\cmidrule(lr){6-7}
  & & & Mean & SD & Mean & SD \\
  \midrule
  logit~1 & $\beta_0$ & 1.00 & 1.02 & 0.22 & 0.95 & 0.23\\
  & $\beta_1$ & 0.70 & 0.72 & 0.23 & 0.65 & 0.20 \\
  & $\rho$ & 0.80 & 0.50 & 0.29 & 0.47 & 0.27 \\
  & $\nu$ & 2.00 & 2.33 & 6.03 & 4.24 & 2.49 \\
  \vspace{3mm}
  & LPML &  & $-$92.30 & 5.59 & $-$91.77 & 5.77 \\
  logit~2 & $\beta_0$ & 1.00 & 1.03 & 0.22 & 0.97 & 0.23\\
  & $\beta_1$ & 0.70 & 0.73 & 0.22 & 0.66 & 0.20 \\
  & $\rho$ & 0.80 & 0.50 & 0.29 & 0.46 & 0.27 \\
  & $\nu$ & 1.00 & 1.56 & 3.57 & 3.79 & 2.52 \\
  \vspace{3mm}
  & LPML &  & $-$91.67 & 5.53 & $-$91.31 & 5.58 \\
  beta-logit & $\beta_0$ & 1.08 & 0.22 & 0.18 & 1.01 & 0.27 \\
  & $\beta_1$ & 0.70 & 0.78 & 0.23 & 0.68 & 0.20 \\
  & $\rho$ & 0.80 & 0.52 & 0.29 & 0.56 & 0.27 \\
  & $\nu$ &  2.00 & 1.10 & 1.09 & 3.99 & 2.37 \\
  & LPML &  & $-$96.16 & 6.27 & $-$88.16 & 8.21 \\
  \bottomrule
\end{tabular}
\end{table}

Similar to the results from Section~\ref{s:po:ss}, when the model
was specified correctly, the true values of regression coefficients
are recovered very well; the dispersion parameter estimate
tended to be bigger than true value; and the dependence parameter
estimate appeared to be downward biased.
When the true model was the beta-logit model, the average LPML value
of the beta-logit model was 8 higher than that of the logit model.
When the true model was the logit~1 or logit~2, however, the average
LPML value of the beta-logit model was very close to (actually
slightly higher than) that of the logit model in both cases.
This implies that the beta-logit model is quite accommodating
and can provide close approximation to the logit model;
with the sample size in our simulation, they are hard to distinguish.

\begin{table}
\caption{Frequencies of model selection using the LPML statistics for the 100 simulated datasets.}
\label{t:bin:cpo}
\centering
\begin{tabular}{ccc}
  \toprule
 True model & \multicolumn{2}{c}{Frequency Selected}\\
 \cmidrule(lr){2-3}
 & logit & beta-logit \\
  \midrule
  logit 1 & 46 & 54 \\
  logit 2 & 49 & 51 \\
  beta-logit & 16 & 84 \\
  \bottomrule
\end{tabular}
\end{table}

Table~\ref{t:bin:cpo} summarizes the frequencies of the models selected
with the highest LPML from all 100 datasets generated under each scenario.
When the true model was the beta-logit model, the LPML criterion
worked effectively, correctly selecting the true model 84 times.
When the true model was logit~1 or logit~2, however, the logit model
and the beta-logit model were selected with almost equal frequency,
indicating that the beta-logit model provides very good approximation
of the logit model with our sample size.

Box plots of the difference in LPML between the correct model
and the misspecified model are shown in Figure~\ref{fig:bern}.
The boxes are surprisingly tight around zero when the true model
is the logit model, indicating that the beta-logit model
approximates the logit model very closely in terms of LPML.
When the true model was the beta-logit model, however, the
LPML value of the logit model was very unlikely to be higher
than that of the correctly specified model.
The majority of all box plots were well above $-5$.
A difference of 4.2 between the two models as observed in
the analysis of presence/absence of \emph{G. nigrolineata}
seems to be quite strong evidence in favor of the beta-logit model.

\section{Discussion}
\label{s:conc}

In geostatistics, the trans-Gaussian kriging approach if often used to 
transform the responses to achieve joint normality \citep{cressie}.
Although the dependence structure in a trans-Gaussian kriging approach
is also a Gaussian copula, our approach is different in several aspects.
Our transformation is not to Gaussian but from Gaussian, and our
model is directly built for the variable of interest, rather than
on some power transformation of it, which may be hard to interpret.
Even when viewed as a to-Gaussian transformation, our 
transformation is margin specific and can incorporate covariates.
From a hierarchical model point of view, our random fields are mostly
useful for model parameters such as Poisson intensity or Bernoulli rate,
which is a different domain than kriging or spatial interpolation.
\citep[see,][]{DeOliveira:etal:1997,Azzalini:Capitanio:1999}.

The proposed models are highly likely to be favored by the
LPML model selection criterion when they are the true models.
Even when they are misspecified, they may still be competitive
by providing a close approximation in the misspecified class 
for small to moderate sample sizes in practice.
Our simulation study for the spatial Poisson regression examined
the performances of three TGMRF models with different marginal
distributions and different parameterizations to incorporate covariates.
The GSC model appears to be more versatile than the traditional LN 
model in that, even when the latter is the true model, the former 
may provide a very close approximation under practical sample size.
The gamma shape model provides another way to improve data fitting.
For the abundance date of \emph{N. tridens}, the GSH model
provided the best fit and sheds light on important predictors.
In the simulation of the spatial Bernoulli regression, the beta-logit
model appeared to be as good as the logit model in terms LPML even when
the data were generated by the logit model, but the opposite was not true.
For the presence of \emph{G. nigrolineata}, the beta-logit
improved the fitting with narrower HPD credible intervals.
In real world applications, where the true model is unknown,
the class of our proposed models may be useful in approximating
the unknown truth.

\section*{Acknowledgments}
This research was partially supported by a Multidisciplinary Environmental Research Award for Graduate Students to M. O. Prates from the Center for Environmental Sciences and Engineering at the University of Connecticut.
M. O. Prates also acknowledges FAPEMIG for partial financial support.
In addition, this research was facilitated by grant numbers BSR-8811902, DEB-9411973, DEB-0080538, and DEB-0218039 from the National Science Foundation to the Institute of Tropical Ecosystem Studies, University of Puerto Rico, and the International Institute of Tropical Forestry as part of the Long-Term Ecological Research Program in the Luquillo Experimental Forest.  Additional support was provided by the USDA Forest Service, the University of Puerto Rico, the Department of Biological Sciences at Texas Tech University, and the Center for Environmental Sciences and Engineering at the University of Connecticut.  The staff of El Verde Field Station provided valuable logistical support in Puerto Rico.  Finally, we thank the mid-sized army of students and colleagues who have assisted with collection of field data over the years.

\bibliographystyle{asa}
\bibliography{tgmrf}

\end{document}